\documentclass[onecolumn]{aastex63}
\usepackage[figuresright]{rotating}
\usepackage{graphicx}
\usepackage{multirow}
\DeclareGraphicsExtensions{.pdf,.jpg,.png}
\usepackage{float}
\usepackage{CJKutf8}

\usepackage{threeparttable}

\begin{document}


\title{Very Large Array Multi-band Radio Imaging of the Triple AGN Candidate SDSS J0849+1114}
\email{sijiapeng@smail.nju.edu.cn; lizy@nju.edu.cn; xinliuxl@illinois.edu}


\author[0000-0001-8492-892X]{Sijia Peng}
\affiliation{School of Astronomy and Space Science, Nanjing University, Nanjing 210023, China}
\affiliation{Key Laboratory of Modern Astronomy and Astrophysics, Nanjing University, Nanjing 210023, China}

\author[0000-0003-0355-6437]{Zhiyuan Li}
\affiliation{School of Astronomy and Space Science, Nanjing University, Nanjing 210023, China}
\affiliation{Key Laboratory of Modern Astronomy and Astrophysics, Nanjing University, Nanjing 210023, China}

\author[0000-0003-0049-5210]{Xin Liu}
\affiliation{Department of Astronomy, University of Illinois at Urbana-Champaign, Urbana, IL 61801, USA}
\affiliation{National Center for Supercomputing Applications, University of Illinois at Urbana-Champaign, 605 East Springfield Avenue, Champaign, IL 61820, USA}

\author[0000-0003-1991-370X]{Kristina Nyland}
\affiliation{National Research Council, resident at the U.S. Naval Research Laboratory, 4555 Overlook Avenue SW, Washington, DC 20375, USA}

\author[0000-0001-9720-7398]{Joan M. Wrobel}
\affiliation{National Radio Astronomy Observatory, P.O. Box 0, Socorro, NM 87801, USA}

\author[0000-0001-9062-8309]{Meicun Hou}
\affiliation{Kavli Institute for Astronomy and Astrophysics, Peking University, Beijing 100871, China}

\begin{abstract}
Kpc-scale triple active galactic nuclei (AGNs), potential precursors of gravitationally-bound triple massive black holes (MBHs), are rarely seen objects and believed to play an important role in the evolution of MBHs and their host galaxies. In this work we present a multi-band (3.0, 6.0 10.0, and 15.0 GHz), high-resolution radio imaging of the triple AGN candidate, SDSS J0849+1114, using the Very Large Array. Two of the three nuclei (A and C) are detected at 3.0, 6.0, and 15 GHz for the first time, both exhibiting a steep spectrum over 3--15 GHz (with a spectral index $-0.90 \pm 0.05$ and $-1.03 \pm 0.04$) consistent with a synchrotron origin. Nucleus A, the strongest nucleus among the three, shows a double-sided jet, with the jet orientation changing by $\sim20\arcdeg$ between its inner  1$''$ and the outer 5\farcs5 (8.1 kpc) components, which may be explained as the MBH's angular momentum having been altered by merger-enhanced accretion. Nucleus C also shows a two-sided jet, with the western jet inflating into a radio lobe with an extent of 1\farcs5 (2.2 kpc). The internal energy of the radio lobe is estimated to be $\rm 5.0 \times 10^{55}$ erg, for an equipartition magnetic field strength of $\rm \sim 160\ \mu G$. No significant radio emission is detected at all four frequencies for nucleus B, yielding an upper limit of 15, 15, 15, and 18 $\rm \mu Jy\ beam^{-1}$ at 3.0, 6.0, 10.0, and 15.0 GHz, based on which we constrain the star formation rate in nucleus B to be $\lesssim 0.4~\rm M_{\odot}~yr^{-1}$. 

\end{abstract}




\section{Introduction} 
Galaxy mergers are central to the widely accepted hierarchical structure formation paradigm \citep{toomre72}. Because most massive galaxies harbor a massive black hole \citep[MBH;][]{Kormendy1995}, mergers should produce binary MBHs \citep{begelman80}. In rarer situations, triple MBHs may form when a subsequent merger with a third galaxy occurs before the first two BHs coalesce \citep{valtonen96}. Therefore, direct evidence for triple MBHs would offer a unique verification of the hierarchical merger paradigm. 

Furthermore, triple MBHs are of significant merit for both understanding key aspects of galaxy formation and probing fundamental physics. For example, triple MBHs are predicted to play a crucial role in regulating the formation of the stellar cores in massive elliptical galaxies \citep{DEGN}. Simulations suggest that triple MBHs scour out stellar cores with mass deficits one or two times the total BH mass, and sizes that are larger than those formed around binary MBHs; this process may be responsible for the unusually large cores observed in some massive elliptical galaxies such as M87 \citep{hoffman07}. In addition, close triple MBHs offer a laboratory for studying the dynamics of three-body interactions in general relativity \citep{DEGN}. Theory suggests that hierarchical systems of close triple MBHs may exhibit phases of very high eccentricity in the inner binary. These systems are thought to produce intense bursts of gravitational waves which could be detectable with ongoing and future low-frequency gravitational wave experiments \citep{amaro10}. The precursors of close triple MBHs, including those that are still separated by $\lesssim$ a few kpc, are of interest, because they may be used to inform the subsequent evolution at closer separations \citep{DEGN}.

Despite the intense theoretical interest, and strong reasons to believe that they exist, triple MBHs that are in direct gravitational interaction have not been conclusively detected. This lack arises mostly because the typical separation of close triples ($\lesssim$ a few parsecs) is too small to resolve at cosmological distances. Kpc-scale triple MBHs are a possible precursor of close triples. They are not yet in direct gravitational interaction, but their separations are both large enough to resolve with current facilities and small enough to be dynamically interesting. Moreover, the typical orbital decay timescale of the kpc phase is long enough ($\gtrsim10^8$\,yr) that it may provide the best chance for direct detection.

Kpc-scale triple MBHs can be identified as kpc-scale triple Active Galactic Nuclei (AGNs), when all three BHs are accreting -- a process expected in gas-rich mergers \citep{hernquist89}. There are two candidate physical triple quasars known, QQ 1429.008 at $z=2.1$ \citep{djorgovski07}, and QQQ J1519+0627 at $z=1.51$ \citep{Farina2013}, but the projected separations between the quasars are much larger, i.e., of 30--50 kpc and of 200 kpc, respectively, and it is unclear whether the host galaxies are in direct merging process, due to the lack of tidal features indicative of ongoing interactions. Until recently only one possible candidate was serendipitously discovered in NGC 3341 \citep{Barth2008}; the disk galaxy contains three nuclei (with two offset nuclei at projected separations of 5.1 and 8.4 kpc from its primary nucleus), which are optically classified as a Seyfert, a LINER, and a LINER or LINER-H {\tiny II} composite, respectively. There is a possible candidate reported at $z=1.35$ \citep{schawinski11} based on indirect evidence from rest-frame optical diagnostic line ratios, although the resolution of the HST grism spectrum is too low to cleanly separate the individual components and it is more likely to be a clumpy star-forming galaxy at $z>1$ rather than a bona-fide triple AGN. There was another candidate triple AGN system (J1502+1115) at $z=0.39$ discovered by the NSF’s Karl G. Jansky Very Large Array (VLA) \citep{Deane2014}, although two nuclei were later found to be radio hot spots instead of a pair of MBHs \citep{Wrobel2014}. A systematic SDSS search found a candidate, SDSS J1027+1749 \citep{Liu2011}, but only one of the nuclei is optically classified as a Seyfert whereas the other two nuclei are a LINER and an AGN/H {\tiny II} composite. Another candidate through a systematic SDSS search is SDSS J1056+5516 \citep{2017ApJ...851L..15K} at $z$ = 0.256, but it is inconclusive since one of the nuclei is a star formation/LINER composite which requires further X-ray or radio observations to confirm. NGC 6240 has also been suggested to host three nuclei of which two are active \citep{Kollatschny2020}.

In this work, we present deep VLA multi-band radio imaging for SDSS J0849+1114 (at $z=0.078$), the best kpc-scale triple AGN candidate known to date \citep{Liu2019,Pfeifle2019a,Foord2021,Foord2021a}. The target was originally identified from a systematic survey of AGN pairs in the optical \citep{Liu2011a} combined with comprehensive follow-ups \citep{Liu2019}. It was also independently identified \citep{Pfeifle2019a} from a systematic search of IR-selected mergers \citep{Pfeifle2019} with the Wide-Field Infrared Survey Explorer all-sky survey \citep{Wright2010}. All of its three nuclei are optically classified as Type 2 Seyferts, where the nuclear emission is obscured in the optical, and their AGN nature was indirectly inferred based on diagnostic narrow emission-line ratios using the BPT diagram \citep{bpt,veilleux87}. The excitation mechanisms estimated from optical diagnostics were inconclusive, because of the often present star formation, dust/gas extinction, and/or shock heating, which may either dilute or mimic AGN excitation. Furthermore, there could be only one or two active MBHs that are ionizing all three galaxies, producing three AGN-like nuclei in the optical. The spatial distribution of ionization parameters estimated from optical emission-line measurements was unable to pin down the locations of the ionizing sources, due both to the proximity of the nuclei and in particular to systematic uncertainties in the electron density measurements \citep{Liu2010b}. While the pilot program successfully measured the radio flux densities for two of its three nuclei and put an upper limit on the third one \citep{Liu2019}, the shallow, single-band data was insufficient to unambiguously confirm the AGN nature in the radio in all but one nucleus. 

Radio detection may provide the most direct evidence for nuclear activity \citep{ho08}. Although the spatial resolutions and/or sensitivities of previously existing FIRST and VLASS images were insufficient, the VLA's sub-arcsecond spatial resolution in its most extended A configuration, together with its superb capability of spectral imaging in the radio, makes it the ideal match for our goals to detect all three AGNs in the radio or impose stringent upper limits in the case of a non-detection. Indeed, high-resolution VLA imaging has been instrumental in confirming many of the currently known handful cases of kpc-scale dual AGNs \citep{Wrobel2014a,Wrobel2014,Fu2015,Fu2015a}. To pin down the nature of the ionizing sources in all three nuclei, we have obtained deeper VLA multi-band radio imaging. Along with the existing complementary multi-wavelength observations from the {\it Hubble Space Telescope} (HST) and {\it Chandra X-ray Observatory}, the new radio observations provide the most sensitive test of the triple AGN hypothesis for SDSS J0849+1114. Along with supplementary data, the much deeper, multi-band VLA observations allow us to confirm the AGN nature with radio spectral measurements in two of the three Seyfert nuclei and set the most stringent upper limit on the radio flux density from the faintest nucleus. We use radio variability as a complementary method of AGN confirmation \citep{Nyland2020}, although no significant radio variability is found. Finally, we present a detailed investigation of the two radio jets associated with the two radio-detected nuclei and address their connection to merger-driven AGN activity \citep{Wrobel2014}.

Throughout this paper, quoted errors are of 1-$\sigma$ unless otherwise specified. We adopt a luminosity distance of 355.5 Mpc and an angular diameter distance of 306.0 Mpc for SDSS J0849+1114, assuming $\Omega_{\rm m}=0.286$, $\Omega_{\Lambda} = 0.714 $, and $\rm H_{0}= 69.6\ km \ s^{-1}\ Mpc^{-1}$.

\section{New VLA Observations and Data Preparation} \label{sec:data}

Observations of SDSS J0849+1114 were performed with the VLA in its B-configuration in April 2019 and A-configuration between August--October 2019 (Program ID: 19A-085; PI: X. Liu).
The B-configuration observations were taken in the X band (central frequency 10.0 GHz) on four epochs for a total on-source integration of 3.38 hr, while the A-configuration observations were taken in the S, C, X, and Ku bands (central frequency 3.0, 6.0, 10.0 and 15.0 GHz) for a total on-source integration of 5.70, 0.87, 1.82 and 1.52 hr, respectively. 
Among these, the S and X bands were observed for two epochs, while the C and Ku bands were observed for only one epoch. 
A phase-referenced mode was adopted with J0842+1835 serving as a phase calibrator. J0319+4130 was used as a leakage calibrator and 3C286 was utilized as a flux density and bandpass calibrator. Table \ref{tab:obsinfor} summarizes the basic observational information.

The raw data were calibrated with the Common Astronomy Software Applications package (CASA, version: 5.6.2) by running the VLA pipeline\footnote{https://casaguides.nrao.edu/index.php?title=VLA\_CASA\_Pipeline-CASA5.6.2} \citep{2007ASPC..376..127M} with additional manual flagging and phase-only self-calibration on the visibilities.
We have examined the X-band images of the six individual epochs for potential flux density variability in either nucleus A or nucleus C (Table \ref{tab:X-band}). No significant ($\rm \lesssim 10\%$) flux density variation was found among either the A-configuration epochs or the B-configuration epochs; a $\sim$20\% difference was found between the mean peak flux density in the A- and B-configuration images, which can be understood as due to the extended nature of both nuclei (see Section~\ref{sec:result}). Therefore, the visibilities of a given band at the same configuration were concatenated using the task `CONCAT'. 
Finally, we set `MTMFS' deconvolve mode in the task `TCLEAN' to obtain the cleaned image. 
The resultant A-configuration concatenated images have a synthesized beam of $\rm 0\farcs63 \times 0\farcs53$, $\rm 0\farcs33 \times 0\farcs27$, $\rm 0\farcs27 \times 0\farcs18$, and $0\farcs21 \times 0\farcs12$, and an RMS of 5, 5, 7, and 6 $\rm \mu Jy\ beam^{-1}$, at the S, C, X, and Ku bands, respectively.
The B-configuration X-band concatenated image has a synthesized beam of $\rm 0\farcs66 \times 0\farcs56$ and an RMS of 5 $\rm \mu Jy\ beam^{-1}$.

\begin{figure}[h]
\centering
\includegraphics[width=0.89\textwidth]{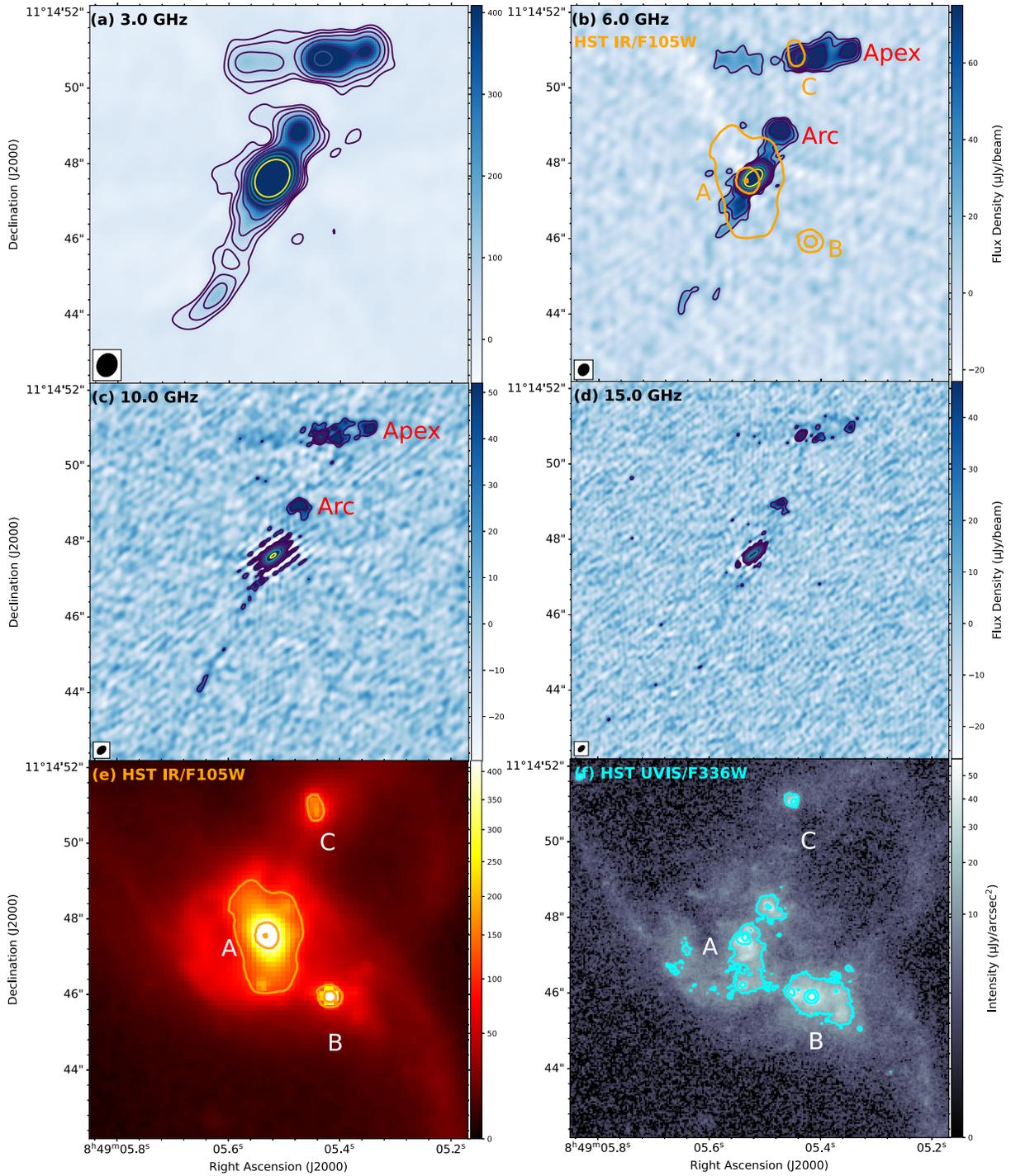}
\caption{ VLA and {\it HST} images of the triple-AGN system SDSS J0849+1114. (a)--(d): radio images at a central frequency of 3.0, 6.0, 10.0 and 15.0 GHz. The radio synthesized beam is indicated in the lower-left corner in panels (a)--(d). The purple to yellow contours are at levels of $\rm (4, 8, 16, 32, 64, 128,256, 512) \times RMS$, which is 5, 5, 7 and 6 $\rm \mu Jy\ beam^{-1}$ at 3, 6, 10, and 15 GHz. In panel (b), $HST$ Y-band intensity contours (orange) are overlaid to indicate the optical positions of nuclei A, B, and C. Two radio features, the Arc and the Apex, are also marked. In panel (c), some weak features along southeast-northwest around nucleus A are artifacts probably due to imperfect self-calibration correction, which do not significantly affect our analysis. (e): {\it HST}/WFC3 IR/F105W (Y-band) image.
The intensity contours are at levels of $\rm (106.8,  427.3, 1709.2)\ \mu Jy\ arcsec^{-2}$.
(f): {\it HST}/WFC3 UVIS/F336W (U-band) image. 
The intensity contours are at levels of $\rm (11.6, 46.4, 185.5)\ \mu Jy\ arcsec^{-2}$.
\label{fig:radio}}  
\end{figure}

We supplement the VLA images with the {\it HST}/WFC3 U-band and Y-band images (Figure \ref{fig:radio}e,f), originally presented in \citet{Liu2019}. 
The {\it HST} images have been registered to the astrometry of SDSS and have an absolute uncertainty of 0\farcs15 \citep{Liu2019}.

\section{Results} 
\label{sec:result}
The VLA images of a $\rm 10''\times 10''$ (corresponding to $\rm 14.8\ kpc \times 14.8\ kpc$) region around SDSS J0849+1114 at 3.0, 6.0, 10.0, and 15.0 GHz are shown in Figure \ref{fig:radio}, which provides new insights to this triple AGN candidate.  
Here the 10.0 GHz image includes only data taken under A-configuration for an optimal angular resolution.

\subsection{The Triple Nuclei}
Both nucleus A and nucleus C are clearly detected in all four bands. \citet{Liu2019} reported 9.0 GHz (X-band) detection of nucleus A and C, and here we have new detections at 3.0, 6.0, and 15.0 GHz.
Moreover, interesting structures are revealed in both nuclei.

\begin{figure}[t]
\centering
\includegraphics[width=0.9\textwidth]{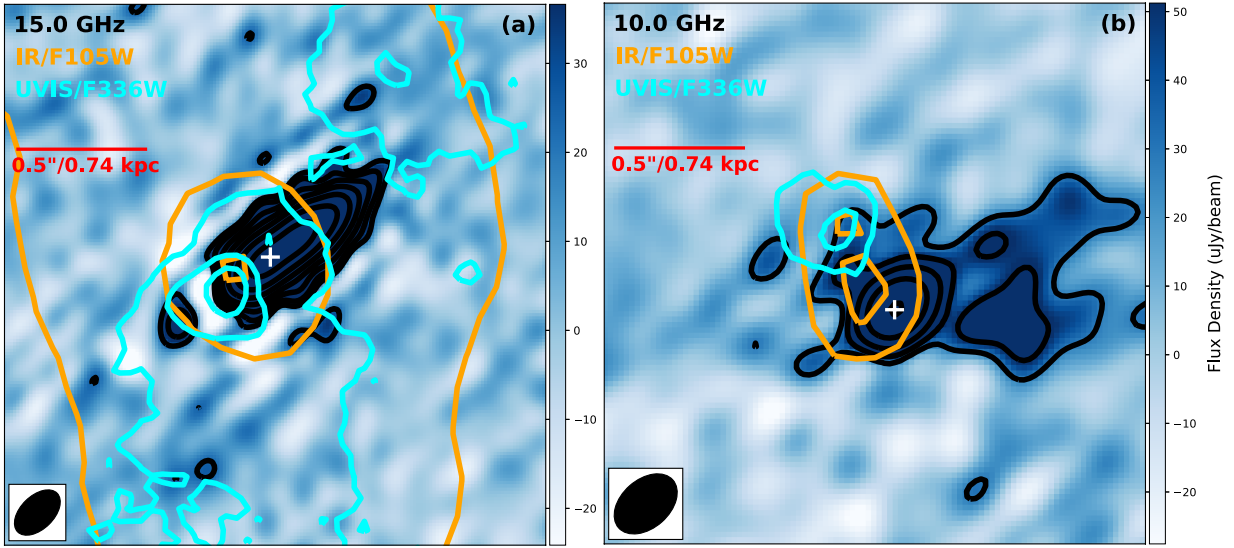}
\caption{A zoom-in view of (a) nucleus A at 15.0 GHz and (b) nucleus C at 10.0 GHz. Each panel has a size of $2\arcsec \times 2\arcsec$. The orange and cyan intensity contours are of the {\it HST} Y-band and U-band, respectively (same as in Figure~\ref{fig:radio}). The radio intensity peak of the nucleus is marked by a white `+' sign. \label{fig:zoomin} }
\end{figure}

Figure \ref{fig:zoomin} provides a close-up view of nucleus A at 15.0 GHz and nucleus C at 10.0 GHz. 
In Figure \ref{fig:zoomin}a, 
nucleus A exhibits an elongated morphology, reminiscent of a radio jet pointing along southeast-northwest at a position angle of $\sim$307$\arcdeg$ (east from the north) with an apparent extent of $\sim$0\farcs6 ($\sim$0.9 kpc).
This elongated feature (larger than the corresponding synthesized beam size) is also evident in the 10.0 GHz image of nucleus A (Figure~\ref{fig:radio}c).
The intensity peak of nucleus A in the {\it HST} Y-band (orange contours; Figure \ref{fig:radio}e) and U-band images (cyan contours; Figure \ref{fig:radio}f) is consistent with the 15.0 GHz intensity peak at [RA, DEC]=$[08^{h}49^{m}5\fs526, +11\arcdeg14\arcmin47\farcs57]$ (marked as the white cross in Figure~\ref{fig:zoomin}a) to within the astrometric uncertainty of $\sim$0\farcs15 between the optical and radio images, suggesting that the latter may be the position of the jet base. 
If this were the case, a discrete knot seen at $\sim$0\farcs3 southeast of the jet base may be the trace of an otherwise insignificant counter-jet, likely due to the relativistic Doppler dimming effect.  
Due to the extended nature of nucleus A, we report its integrated flux density at each frequency (Table \ref{tab:nuclei}), which is measured 
using the CASA task IMFIT with a two-dimensional Gaussian model.
The resultant integrated flux density ($S_\nu$) is $\rm 17.75 \pm 0.56 $, $\rm 10.28\pm 0.56 $, $\rm 6.26\pm 0.21$ and $\rm 4.21\pm 0.17$ $\rm  m Jy$ at  3, 6, 10, and 15 GHz, respectively. Using orthogonal distance regression, we derive a spectral index of $\alpha = -0.90 \pm 0.04$ ($S_\nu \propto \nu^{\alpha}$) for nucleus A. Accounting for the coherence loss, the measurements of nucleus A from the VLA agree with the results from European VLBI Network (EVN) at 1.7 GHz \citep{2019A&A...630L...5G}.

Figure \ref{fig:zoomin}b shows that the 10 GHz emission from nucleus C is dominated by a compact core, which peaks at [RA, DEC]=$[08^{h}49^{m}05\fs436, +11\arcdeg14\arcmin50\farcs76]$ 
and is coincident with the southern intensity peak in the Y-band image (orange contours; Figure \ref{fig:radio}e), to within the astrometric uncertainty of 0\farcs15.
On the other hand, the U-band intensity peak (also coincident with the northern peak in the Y-band) is significantly offset ($\sim0\farcs5$ to the northeast) from the radio peak.
If the latter marked the true position of the putative MBH, the U-band peak may be understood as the site of circumnuclear star formation. Significant 10.0 GHz emission seen to the immediate northeast of the radio core may be associated with this star-forming activity. 
Strong extended emission is also present to the west of the radio core, exhibiting an edge-brightened morphology reminiscent of a radio bubble. This bubble-like feature may be connected to the radio core with weak emission. 
The integrated flux density of nucleus C is measured to be $\rm 1.62\pm 0.09$, $\rm 0.83\pm 0.07$,$\rm 0.45\pm 0.04$ and $\rm 0.32\pm 0.02$ $\rm mJy$ at 3.0, 6.0, 10.0, and 15.0 GHz, respectively. The best-fit spectral index is $\rm -1.03 \pm 0.04$. 

Despite the unprecedented sensitivity of these images, nucleus B remains undetected in all four bands. Therefore, we estimate a 3-$\sigma$ upper limit for the peak flux density of nucleus B, which is 15, 15, 15, and 18 $\rm \mu Jy\ beam^{-1}$ at 3.0, 6.0, 10.0, and 15.0 GHz, respectively. Here the 10 GHz upper limit is from the B-configuration concatenated image, which has an RMS 1.4 times lower than that of the A-configuration image. These limits are much tighter than reported by  \citet{2019A&A...630L...5G} based on the EVN observations, which is $\rm < 450\ \mu Jy\ beam^{-1}$ at 1.7 GHz.
Figure \ref{fig:sed} displays the observed radio spectral energy distributions of the triple nucleus (upper limits in the case of nucleus B), along with the best-fit power-law.

\subsection{Extended Features}

\begin{figure}[h]
\centering
\includegraphics[width=1\textwidth]{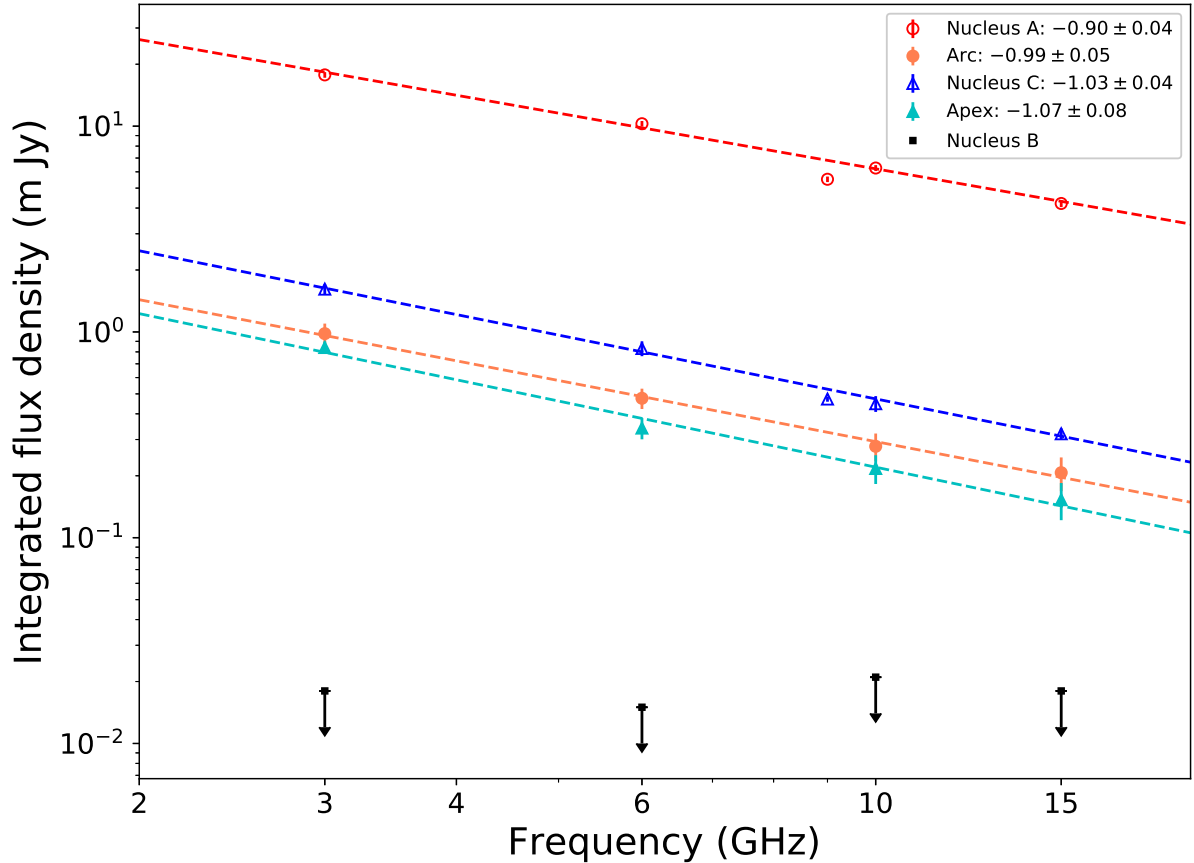}
\caption{The radio spectral energy distributions of nucleus A (red open circle), nucleus C (blue open triangle), the Arc (orange filled circle) and the Apex (cyan filled triangle). The integrated flux density of nucleus A and C from \citet{Liu2019} at 9 GHz are also plotted. 1-$\sigma$ errors are plotted, but are too small to be discerned for nucleus A and nucleus C. The dashed lines are the best-fitted power-law, with values of the spectral index given in the insert. 3-$\sigma$ upper limits of nucleus B are also shown (black arrows). \label{fig:sed}  } 
\end{figure}

The VLA images also reveal remarkable features on the galactic scales.
Most prominent is a linear feature through nucleus A in the 3.0, 6.0, and 10.0 GHz images (Figure~\ref{fig:radio}), which is reminiscent of a two-sided jet. To distinguish it from the jet-like feature within half arcsec of nucleus A (Figure~\ref{fig:zoomin}a), we shall refer to this linear feature as the outer jet (and the former as the inner jet). 
Remarkably, the position angle of the outer jet is $\sim$327$\arcdeg$, i.e., deviating from that of the inner jet by 20$\arcdeg$.
The outer jet is more extended on the southeast side, reaching an apparent extent of at least $\sim5\farcs5$ ($\sim$8.1 kpc) in the 3.0 GHz image, which is most sensitive to extended features.  
The S-band image shows that the far side is brightened and slightly bent eastward, perhaps due to deflection by some external pressure. 
On the northwestern side, the far end of the outer jet is defined by an arc-shaped feature located at $\sim$1\farcs8 from nucleus A, which is clearly seen in the 10.0 and 15.0 GHz images and reminiscent of a jet-driven shock. That the `arc' and nucleus A are connected by weak emission is evident in the 3.0 and 6.0 GHz images. 
The markedly different radio morphology on the two sides of nucleus A might be due to an intrinsic difference in the galactic-scale environment or the relativistic Doppler dimming.
We measure the integrated flux densities of the arc in individual bands adopting a two-dimensional Gaussian model, which are given in Table \ref{tab:nuclei} and plotted in Figure~\ref{fig:sed}. The best-fit spectral index of the arc is found to be $-0.99 \pm 0.05$, consistent with synchrotron radiation from shock-accelerated high-energy particles.
The arc has no clear counterpart in the {\it HST} images.
A visual examination of the {\it Chandra} image of \citet[][Fig.6 therein]{Liu2019} (see also \citealp{Pfeifle2019a}) suggests that a few X-ray photons are detected at the position of the arc, but deeper {\it Chandra} exposures are necessary to confirm this marginal excess.
We have also examined the optical spectra presented in \citet{Liu2019} for potential shock-induced spectral features at the position of the arc. No clear evidence of this kind is found.

A two-sided linear feature through nucleus C is also evident in all four bands but most clearly seen in the 3.0 and 6.0 GHz images (Figure~\ref{fig:radio}). This linear feature has a length of $\sim$1.5\arcsec\ ($\sim$2.2 kpc) on both sides and a width of $\sim$ 0.6\arcsec\ ($\sim$0.9 kpc). 
On the western side, the linear feature ends at a prominent knot (hereafter called the `apex'), which may be the top of the bubble-like feature shown in Figure~\ref{fig:zoomin}b.  
Emission on the eastern side is substantially fainter compared to its western counterpart and shows no sign of a jet-inflated bubble.
Again, this difference on the two sides of nucleus C might be due to an intrinsic difference in the environment.
We find no significant optical or X-ray counterpart of this two-sided radio feature.
We measure the integrated flux densities of the apex in individual bands adopting a two-dimensional Gaussian model (Table \ref{tab:nuclei} and Figure \ref{fig:sed}). The best-fit spectral index is $ \rm -1.07 \pm 0.08$, again suggestive of a synchrotron origin.

\section{Discussion and Conclusion}
\subsection{Extended Jets from Nucleus A}

The inner jet in nucleus A points along southeast-northwest at a position angle of $\sim 307\arcdeg$ extending to $\rm 0.6\arcsec$ ($\rm 0.9$ kpc). The inner counter-jet can be resolved at 15.0 GHz, which follows the same direction. On the scale up to $1\arcsec$ ($\rm 1.5$ kpc) away from the peak of nucleus A, the outer jets on both sides deviate from the position angle of the inner jet by $\sim 20\arcdeg$. This indicates that the jet and the counter-jet have both turned an angle of $20\arcdeg$ on their way forward.
However, it is too coincidental to explain the same offset angle of the jets on both sides due primarily to an external pressure. We speculate that a more natural scenario to explain the offset could be that a spinning black hole is powering episodic jets. An older jet points at a certain position angle. When a newly born younger jet emerges, the jet axis has rotated by another certain angle, resulting in the younger jet not moving forward at the previous position angle of the older jet. While galaxy mergers are not a necessary condition for radio jet reorientation, it is plausible that the ongoing galaxy merger may have promoted more gas inflows that could affect both the accretion states of the active MBHs and the dynamical state of the circumnuclear medium.  
Indeed, theoretical studies demonstrate that the spin axis of a strongly accreting MBH can be significantly altered and ultimately aligned with the angular momentum of a gas inflow feeding the accretion disk over a timescale of $\lesssim$ Myr \citep{2018MNRAS.477.3807F, 2021MNRAS.500.3719C}.

We estimate the minimum time gap between the old and young jets of nucleus A. By assuming the jet has an average speed of $ 0.1c$ ($c$ is the speed of light), it takes $\sim \rm 1.5 \times 10^{5}$ yr to travel 4.7 kpc. As for the young jet, a smaller distance of 0.9 kpc needs $\sim \rm 2.9 \times 10^{4}$ yr of travel. Therefore, the minimum time gap is around $\rm 1.2 \times 10^{5}$ yr, compatible with the aforementioned timescale suggested by theoretical studies \citep{2018MNRAS.477.3807F, 2021MNRAS.500.3719C}.
We note that the relatively strong magnetic field could explain a negative spectral index. A rapid synchrotron cooling happens in a strong magnetic field.

\subsection{Nature of the Western Lobe in Nucleus C}

The 15 GHz (Ku-band) image well describes the outline of the western lobe that has a length of $\sim$1.5\arcsec\ ($\sim$2.2 kpc) and a width of $\sim$ 0.6\arcsec\ ($\sim$0.9 kpc). The Apex seems to be the hot spot of the western lobe. Approximating the lobe as an ellipsoid with a circular cross-section and assuming that the line-of-sight is close to edge-on, the volume of the lobe is $\sim$ 0.8 $\rm kpc^{3}$. The integrated flux density in such a region is $\rm 0.81\pm 0.03$, $\rm 0.52\pm 0.04$ and $\rm 0.39\pm 0.02$ mJy at 6.0, 10.0 and 15.0 GHz, respectively, and at least 1.35 mJy at 3.0 GHz.
The average radio spectral index of this region is $\rm -0.78 \pm 0.02$. Adopting the classic energy equipartition assumption, the minimum energy density ($u_{\rm min}$, equation 25 in \citet{2004IJMPD..13.1549G}) in the lobe is estimated to be $\sim \rm 2.3\times 10^{-9}\ erg\ cm^{-3}$, for an equipartition magnetic field strength of $B_{\rm eq}= (24\pi u_{\rm min}/7)^{\frac{1}{2}} \approx 160\ \mu$G.
The total energy (i.e., a sum of magnetic and particle energy) of the lobe amounts to $\rm 5.0 \times 10^{55}$ erg, which is comparable to the work done during its expansion.

Because the steep spectrum extends to 3.0 GHz, the synchrotron cooling timescale is at least more than $\rm 4\times 10^{5} $ yr based on the equipartition magnetic field strength. If the jet had a speed of 0.1c, it would need a travel time of $\rm 7\times 10^{4} $ yr. In fact, the lobe expansion could take more time by at least a factor of 10. Here, we choose a minimum timescale of $\rm 7\times 10^{5} $ yr and the jet power is estimated to be $ \rm \sim 4\times 10^{42}\ erg\ s^{-1}$. Such power is plausibly available from nucleus C which is a Seyfert galaxy with an estimated black hole mass of $5 \times 10^{6}\rm~M_{\odot}$ \citep{Liu2019,2015ApJ...813...82R}. 
On the other hand, it is hard to determine any visible star-forming impact from the jet/lobe feedback in such a short timescale. 

The lobe associated with nucleus C is an interesting analog to the famous case of the Circinus galaxy.
Exhibiting two extended edge-brightened radio lobes each with a size of $\sim 1.5$ kpc, the Circinus galaxy has an estimated jet power of $\rm \sim 10^{41}\ erg\ s^{-1}$ \citep{2012ApJ...758...95M}. This is about an order of magnitude lower than the one we find above for the lobe associated with nucleus C, and so is the total energy involved, $\rm 2 \times 10^{55}\ erg $ \citep{2012ApJ...758...95M}.

\subsection{Updated Radio Constraint on the Star Formation Rate in Nucleus B}

The non-detection of nucleus B gives an upper limit of 15, 15, 15, and 18 $ \rm \mu Jy\ beam^{-1}$ at 3, 6, 10, and 15 GHz, respectively. Assuming $\rm S_{\nu} \propto \nu^{-0.59}$, which is appropriate for star-forming galaxies \citep{2018A&A...611A..55K}, 
we extrapolate our 3 GHz measurement to 1.45 GHz, which results in $\rm 28\ \mu Jy\ beam^{-1}$. Using the $L_{\rm 1.4 GHz} - \rm SFR_{UV+TIR}$ correlation \citep{Davies2016,Davies2017}, the inferred $3\sigma$ upper limit of star formation rate (SFR) is $\rm 0.4\ M_{\odot}\ yr^{-1}$. This is slightly more stringent than but is broadly consistent with the $3\sigma$ SFR upper limit of $<0.8\rm~M_{\odot}\ yr^{-1}$ reported in \citet{Liu2019}. A flatter spectral index would lead to a lower SFR, thus the inferred limit is a conservative value.

\citet{Liu2019} also found the dust attenuation corrected SFR is $\rm \sim 0.2 - 7 \ M_{\odot} \ yr^{-1}$ derived from {\it HST} U-band and continuum-index. Based on near-infrared spectroscopy from the Large Binocular Telescope, \citet{Pfeifle2019a} obtained an SFR of $\rm \sim 0.48 \ M_{\odot} \ yr^{-1}$ for nucleus B. Both estimates are consistent with our radio limit. In addition, based on the Kennicutt-Schmidt law \citep{2012ARA&A..50..531K}, our limit is also in agreement with the non-detection of molecular gas in nucleus B at arcsecond resolution, which corresponds to an upper limit of $\rm 79\ M_{\odot}\ pc^{-2}$ in molecular gas surface density (M. Hou et al. in preparation). Therefore, our VLA results reinforce the conclusion that star-forming activity alone is insufficient to account for the observed X-ray flux in nucleus B, and an additional heating source such as AGN and/or shocks are required \citep{Liu2019,Pfeifle2019a}, although such additional power is not necessarily produced by the putative MBH in nucleus B.\\

In summary, we have presented new VLA observations of SDSS J0849+1114 at 3.0, 6.0, 10.0, and 15 GHz, which provide an unprecedented radio view for this triple AGN candidate. Two of the three nuclei, nucleus A and C, are detected for the first time at 3.0, 6.0, and 15 GHz. They both show a steep spectrum over 3--15 GHz, consistent with an origin of synchrotron radiation. The high-resolution images also reveal kpc-scale extended features related to both nuclei, which can be  attributed to AGN-driven jets/outflows. Nucleus B remains undetected at all four frequencies, and further studies are warranted to unambiguously determine the nature of this nucleus.

\acknowledgments 
S.P and Z.L. acknowledge support by the National Key Research and Development Program of China (2017YFA0402703) and National Natural Science Foundation of China (grants 11873028, 11473010). X.L. acknowledges support from NSF grant AST-2108162. M.H. acknowledges support by the fellowship of China National Postdoctoral Program for Innovation Talents (grant BX2021016). We thank Dr. Binbin Zhang for his help with computing resources. This research was supported in part by the National Science Foundation under PHY-1748958.

The National Radio Astronomy Observatory is a facility of the National Science Foundation operated under cooperative agreement by Associated Universities, Inc.

Based in part on observations made with the NASA/ESA Hubble Space Telescope, obtained at the Space Telescope Science Institute, which is operated by the Association of Universities for Research in Astronomy, Inc., under NASA contract NAS 5-26555. These observations are associated with program number GO-12363 (PI: X. Liu).


\clearpage

\begin{deluxetable*}{cccccccc}
\tablecaption{Log of VLA Observations\label{tab:obsinfor}}
\tablecolumns{8}
\tablenum{1}
\tablewidth{0pt}
\tablehead{
\colhead{Band} & \colhead{Frequency} & \colhead{Bandwidth} & \colhead{Configuration} & \colhead{Time} &\colhead{Date} &\colhead{Synthesized Beam} &\colhead{RMS} \\
\colhead{} & \colhead{(GHz)} & \colhead{(GHz)} &\colhead{} &\colhead{(hr)} &\colhead{(2019)}& \colhead{$(\rm '',\ '',\ ^{o})$} & \colhead{($\rm \mu Jy~beam^{-1}$)} 
}
\colnumbers
\startdata
S & 3.0 & 2.0 & A & 5.70  & Oct 4/5 & $\rm 0\farcs63 \times 0\farcs53, -22.4\arcdeg$ &5 \\
C & 6.0  & 4.0& A & 0.87  & Oct 7 & $\rm 0\farcs33 \times 0\farcs27, -33.6\arcdeg$  &5 \\
X & 10.0 & 4.0 & A & 1.82 & Aug 17/25 & $\rm 0\farcs27 \times 0\farcs18, -48.5\arcdeg$ & 7  \\
&  & & B &  3.38  &  Apr 16/25/26/26 & $0\farcs66\times 0\farcs56, -24.2\arcdeg$ & 5 \\ 
Ku & 15.0  & 6.0  & A & 1.52  & Oct 11 & $\rm 0\farcs21 \times 0\farcs12, -45.6\arcdeg$ & 6 \\
\enddata

\tablecomments{(1) Observational band. (2)-(3) Central frequency and Bandwidth. (4) VLA array configuration. The longest and shortest baseline of A-configuration are 36 and 0.68 kilometers, while those of B-configuration are 36 and 0.21 kilometers. (5) The on-source integration time. (6) Date of observation, in the year 2019. (7) Synthesized beam, including the FWHM of the major and minor axes and the position angle. (8) The image RMS level.}
\end{deluxetable*}

\begin{deluxetable*}{cccc}
\tablecaption{10.0 GHz (X-band) Peak Flux Density in Different Epochs \label{tab:X-band}}
\tablecolumns{4}
\tablenum{2}
\tablewidth{5pt}
\tablehead{
\colhead{Date} &\colhead{Configuration} &  \colhead{Nucleus A} & \colhead{Nucleus C} \\
\colhead{(2019)} &\colhead{} & \colhead{($\rm m Jy~beam^{-1}$)} &  \colhead{($\rm m Jy~beam^{-1}$)}
}
\colnumbers
\startdata
April 16 & B & $\rm 5.77 \pm 0.18$  & $\rm 0.53 \pm 0.03$ \\ 
April 25 & B & $\rm 5.32 \pm 0.16$ &  $\rm 0.47 \pm 0.03$ \\
April 25 & B & $\rm 5.69 \pm 0.18$ &  $\rm0.52 \pm 0.02 $ \\
April 26 & B & $\rm 6.03 \pm 0.19$ &  $\rm 0.55 \pm 0.02$ \\
Aug 17 & A & $\rm 4.03 \pm 0.13$ & $\rm 0.44 \pm 0.03$ \\
Aug 18 & A & $\rm 4.00 \pm 0.13$ & $\rm 0.46 \pm 0.03$ 
\enddata

\tablecomments{ (1) The date of the epoch, in 2019. (2) Array configuration. (3)  Peak flux density of nucleus A.  (4) Peak flux density of nucleus C. The errors are at 1-$\sigma$ level including a 3\% relative uncertainty in the flux calibration added in quadrature.}
\end{deluxetable*}

\begin{deluxetable*}{c|cc|cc|cc|cc}
\tablecaption{Multi-band flux density  \label{tab:nuclei}}
\tablecolumns{8}
\tablenum{3}
\tablewidth{0pt}
\tablehead{
\colhead{Name} & \multicolumn{2}{c}{3.0 GHz (S-band)} & \multicolumn{2}{c}{6.0 GHz (C-band)} & \multicolumn{2}{c}{10.0 GHz (X-band)} & \multicolumn{2}{c}{15.0 GHz (Ku-band)}\\
\colhead{} & \colhead{Peak} & \colhead{Integrated} &\colhead{Peak} & \colhead{Integrated} &\colhead{Peak} & \colhead{Integrated} &\colhead{Peak} & \colhead{Integrated} \\
\colhead{} &\colhead{($\rm m Jy~beam^{-1}$)} &\colhead{($\rm m Jy$)}& \colhead{($\rm m Jy~beam^{-1}$)} &\colhead{($\rm m Jy$)} & \colhead{$(\rm \mu Jy~beam^{-1})$} &\colhead{($\rm m Jy$)} & \colhead{$(\rm \mu Jy~beam^{-1})$} &\colhead{($\rm m Jy$)}
}
\colnumbers
\startdata
Nucleus A & $\rm 15.22 \pm 0.47$ & $\rm 17.75 \pm 0.56$ &
$\rm 7.21 \pm 0.22$ & $\rm 10.28 \pm 0.32$ & 
$\rm 4.08 \pm 0.13$ & $\rm 6.26 \pm 0.21 $ & 
$\rm 2.33 \pm 0.08$ &  $\rm 4.21 \pm 0.17$ \\
Nucleus B & $< $ 0.018 & -  & $< $ 0.015 & - & $< $ 0.021 & - & $< $ 0.018 & -\\
Nucleus C & $\rm 0.97 \pm 0.04$ & $\rm 1.62 \pm 0.09 $ 
& $\rm 0.55 \pm 0.03$ &  $\rm 0.83 \pm 0.07 $ 
& $\rm0.45 \pm 0.02 $ &  $\rm 0.45 \pm 0.04$ 
&  $\rm 0.35 \pm 0.01$ &  $\rm 0.32\pm 0.02$ \\
Arc& $\rm 0.52 \pm 0.04$ & $\rm 0.98 \pm 0.12$ 
& $\rm 0.18 \pm 0.02$ & $\rm 0.48 \pm 0.05 $ 
& $\rm0.10 \pm 0.01 $ &  $\rm 0.28 \pm 0.04$ 
&  $\rm 0.055 \pm 0.008$ &  $\rm 0.207 \pm 0.039$ \\
Apex & $\rm 0.47 \pm 0.02$ & $\rm 0.84 \pm 0.06$  
& $\rm 0.15\pm 0.01$ & $\rm 0.34 \pm 0.04 $ 
& $\rm0.08 \pm 0.01 $ &  $\rm 0.22 \pm 0.03$
&  $\rm 0.046 \pm 0.007$ & $\rm 0.153 \pm 0.031$ \\
\enddata

\tablecomments{(1) Object name. (2), (4), (6), and (8) Peak flux density at the four central frequencies. The errors are at 1-$\sigma$ level including a 3\% relative uncertainty in the flux calibration added in quadrature. Upper limits for nucleus B are of 3-$\sigma$. (3), (5), (7), and (9) Integrated flux density at the four central frequencies.}
\end{deluxetable*}




\facilities{HST (WFC3), VLA.}

\clearpage


\bibliography{tripleAGN}{}
\bibliographystyle{aasjournal}



\end{document}